\documentclass[journal]{IEEEtran}
\IEEEoverridecommandlockouts
\usepackage[pdftex]{graphicx}
\graphicspath{{../pdf/}{../jpeg/}{../png/}}
\DeclareGraphicsExtensions{.pdf,.jpeg,.png}

\usepackage[cmex10]{amsmath}
\usepackage{amssymb} 
\usepackage{amsthm}
\usepackage{xspace}
\usepackage{comment}
\usepackage{subcaption}
\usepackage{algorithmic}
\usepackage{algorithm}
\usepackage{amsmath}
\usepackage{cite}
\usepackage{color}
\usepackage{array}
\usepackage{url}
\usepackage{enumerate}
\usepackage{multirow}
\usepackage{enumitem}

\usepackage{url}
\usepackage{todonotes}

\makeatletter
\def\endthebibliography{%
	\def\@noitemerr{\@latex@warning{Empty `thebibliography' environment}}%
	\endlist
}
\makeatother

\IEEEoverridecommandlockouts
\def\be { \begin{eqnarray} }
	\def\ee { \end{eqnarray} }
\def\ba {\boldsymbol{a}}

\def\be { \begin{eqnarray} }
\def\ee { \end{eqnarray} }
\def\ba {\boldsymbol{a}}

\def\bn {\boldsymbol{n}}
\def\bl {\boldsymbol{l}}
\def\bq {\boldsymbol{q}}

\def\bp {\boldsymbol{p}}
\def\bw {\boldsymbol{w}}
\def\bA {\boldsymbol{A}}

\def\bS {\boldsymbol{S}}

\def\bL {\boldsymbol{L}}

\def\bR {\boldsymbol{R}}
\def\bH {\boldsymbol{H}}

\def\bW {\boldsymbol{W}}
\def\bQ {\boldsymbol{Q}}
\def\bW {\boldsymbol{W}}

\def\bth {\boldsymbol{\theta}}

\def\diag {\mathrm{diag}}

\def\diag {\mathrm{diag}}

\def\bphi {\boldsymbol{\phi}}

\def\BibTeX{{\rm B\kern-.05em{\sc i\kern-.025em b}\kern-.08em
    T\kern-.1667em\lower.7ex\hbox{E}\kern-.125emX}}

\ifCLASSINFOpdf
\else
\fi
\begin{document}
%
\title{Min-max Decoding Error Probability Optimization in RIS-Aided Hybrid TDMA-NOMA Networks}

%
%
%

\author{Tra~Huong~Thi~Le,~       
        and~Yan~Kyaw~Tun,~\IEEEmembership{Member,~IEEE}
\thanks{Tra.~H.~T.~Le is with the Department
of Information Technology, FPT University, FPT Da Nang Technology Urban Area, Hoa Hai Ward, Ngu Hanh Son District, Da Nang City, Vietnam, e-mail: tralth4@fe.edu.vn.}
\thanks{Yan Kyaw Tun is with the Department of Electronic Systems, Aalborg University, 2450 København, Denmark, e-mail: ykt@es.aau.dk}.}

%
%

\markboth{Submitted for IEEE Transactions on Network and Service Management}%
{Shell \MakeLowercase{\textit{et al.}}: Bare Demo of IEEEtran.cls for IEEE Journals}
%



\maketitle

\begin{abstract}
One of the primary objectives for future wireless communication networks is to facilitate the provision of ultra-reliable and low-latency communication services while simultaneously ensuring the capability for vast connection. In order to achieve this objective, we examine a hybrid multi-access scheme inside the finite blocklength (FBL) regime. This system combines the benefits of non-orthogonal multiple access (NOMA) and time-division multiple access (TDMA) schemes with the aim of fulfilling the objectives of future wireless communication networks. In addition, a reconfigurable intelligent surface (RIS) is utilized to facilitate the establishment of the uplink transmission between the base station and mobile devices in situations when impediments impede their direct communication linkages. This paper aims to minimize the worst-case decoding-error probability for all mobile users by jointly optimizing power allocation, receiving beamforming, blocklength, RIS reflection, and user pairing. To deal with the coupled variables in the formulated mixed-integer non-convex optimization problem, we decompose it into three sub-problems, namely, 1) decoding order determination problem, 2) joint power allocation, receiving beamforming, RIS reflection, and blocklength optimization problem, and 3) optimal user pairing problem. Then, we provide the sequential convex approximation (SCA) and semidefinite relaxation (SDR)-based algorithms as potential solutions for iteratively addressing the deconstructed first two sub-problems at a fixed random user pairing. In addition, the Hungarian matching approach is employed to address the challenge of optimizing user pairing. In conclusion, we undertake a comprehensive simulation, which reveals the advantageous qualities of the proposed algorithm and its superior performance compared to existing benchmark methods.

\end{abstract}


%
\IEEEpeerreviewmaketitle

\section{Introduction}
Ultra-reliable and low latency communication (URLLC) has recently garnered increased interest because of its potential to support a wide range of essential applications \cite{saad2019vision}. In the context of URLLC, the transmission of small packets, referred to as finite blocklengths (FBL), is commonly employed in order to adhere to stringent reliability and latency criteria. Within the context of FBL (Flexible Block Length) systems, there exists a trade-off between the benefits derived by coding and the associated latency.  Toward this end, the Shannon capacity formula is not inappropriate for FBL. Polyanskiy et al. have derived an approximate equation for the achievable rate with the finite blocklength codes, which is a function of the signal-to-noise (SNR), the blocklength, and the decoding error probability \cite{polyanskiy2010channel}. The findings of the study indicate that the Shannon rate serves as the maximum attainable rate within the finite blocklength regime, taking into account both the chance of decoding errors and the transmission blocklength. Furthermore, (FBL) systems are susceptible to deterioration in transmission rates and inevitable mistakes in packet decoding, particularly in environments with severe propagation characteristics that result in poor channel conditions for users. 

On the other hand, reconfigurable intelligent surface (RIS) technology holds promise for improving URLLC system efficiency by counteracting the negative effects of the wireless channel's random propagation nature \cite{liu2022reconfigurable,wu2021intelligent}. The wireless environment undergoes reconfiguration with the assistance of RIS, resulting in the transformation of random wireless channels into somewhat deterministic beamforming designs. In contrast to conventional relays, the RIS operates without any delay due to the absence of analog-to-digital conversion through radio frequency (RF) chains \cite{li2022exploiting,yaswanth2023energy}. Moreover, if the direct communication link is disrupted, RIS can create a communication link between the transceivers. As a result, in the event of a disruption in the direct communication link, the RIS has the capability to establish a communication link between the transceivers. Consequently, the integration of RIS into the system has the potential to enhance reliability, minimize delay, and support URLLC-based wireless systems. Due to its promising potential, several previous studies have been dedicated to examining the effects of RIS on URLLC. For instance,  in \cite{xie2021user}, the optimization of user grouping, blocklength allocation, and reflective beamforming was performed in a combined manner. The objective of this optimization was to reduce the overall delay experienced by all users while ensuring that their particular dependability needs were met. The works in \cite{hashemi2022deep,ghanem2021joint} explained one alternative performance metric for the performance gain of the RIS-aided URLLC system. The study in  \cite{tehrani2021resource} aimed to reduce the transmit power of base stations (BS) in a secure system that utilizes reconfigurable intelligent surfaces (RIS) to aid in multiple-input-single-output (MISO) URLLC.  This was achieved by developing techniques for constructing artificial noise, BS beamformers, and RIS passive beamforming. The aforementioned studies have provided evidence that the optimization of intelligent reflecting surface (IRS) beamforming can enhance the channel quality of wireless communication networks.

Moreover, offering URLLC to large numbers of users with constrained resources is one of the most difficult issues in 6G compared to 5G. By sharing the bandwidth
among users (and avoiding the requirement to split the blocklength in the process), NOMA may achieve great spectrum efficiency in comparison to regular OMA. This lowers communication latency for the users. Recent research has examined the effectiveness of NOMA in the URLLC system \cite{xie2021user,kotaba2019improving,ren2019joint,chen2019optimal,han2019energy,ahsan2022reliable}. The authors of \cite{xie2021user} put up a very dependable resource allocation strategy to address the NOMA-related dependability issue for combined uplink and downlink transmission in URLLC. A static multiuser NOMA-URLLC framework based on hybrid automated repeat request re-transmissions was created by the authors in \cite{kotaba2019improving}.
To illustrate the functionality of NOMA-URLLC, a range of network setups with persistent user connectivity were simulated inside this framework. Four distinct transmission strategies were examined in a NOMA-based URLLC scenario in \cite{ren2019joint}. While guaranteeing the error probability of the user under the ideal channel condition, the author optimized the user's error probability in the presence of the worst channel condition. NOMA-URLLC systems perform better than OMA-URLLC schemes under all circumstances, as shown by \cite{kotaba2019improving} and \cite{ren2019joint}. The study in \cite{chen2019optimal} examined the ET maximization problem of a multicarrier NOMA system in NOMA-URLLC using a dynamic programming technique. Based on error probabilities and the maximum transmit power, the simulation results showed that NOMA-URLLC systems operate effectively under all given conditions. The fading of wireless channels, which might result in low channel quality, makes it impossible to constantly satisfy the rigorous criteria of dependability and latency in NOMA-URLLC networks. The reflecting amplitudes and phase shifts of an RIS's reflecting elements can be adjusted to fix this problem. Furthermore, RIS-reflecting beamforming optimization can offer extra degrees of freedom to improve the communication performance of URLLC and NOMA networks.

Recently, integrating NOMA and RIS for URLLC system has drawn a lot of interest. Effective beamforming in RIS-aided networks provides 90\% greater rate performance than its orthogonal equivalent, according to the authors of \cite{deshpande2023resource}. Furthermore, using an RIS-aided network for the system under consideration improves rate performance by 40–50\% and 56–80\%, respectively, over using random phase-shifting at the RIS and using a system without an RIS.
\cite{vu2022intelligent} shown that, compared to its orthogonal multiple access counterpart, the optimal phase-shift determination at the RIS in RIS-aided short-packet NOMA systems under perfect and imperfect consecutive interference cancellation may increase block error rate and throughput. Then, \cite{vu2023star} revealed
that the considered system achieves a higher achievable sum
rate than RIS-based short-packet orthogonal multiple access
ones.
Furthermore, the work in \cite{zhang2023throughput} showed that RIS-reflecting beamforming can assist in increasing the significant sum throughput of all users in NOMA-URLLC. In addition, the RIS-NOMA-aided URLLC communication was also introduced into the different scenarios such as the presence of hardware impairments \cite{xia2023ris}, mobile edge computing \cite{yang2022energy} and smart grid \cite{yang2022resource}.

However, the NOMA scheme introduces interference among the users and depends on the execution of successive interference cancellation (SIC) \cite{zhu2022low}. This will significantly
inﬂuence the QoS of the transmissions, e.g., the reliability and the delay due to retransmissions. Based on the findings above, combining the benefits of both NOMA and TDMA is critical to achieve a compromise between longer blocklength and low/no interference to improve massive access while offering URLLC services. 
Recently, researchers have started to study the integration of RIS and hybrid NOMA-TDMA \cite{al2023self,lei2023hybrid,lyu2023hybrid,chen2022active} and the incorporation of hybrid NOMA-TDMA and URLLC\cite{zhu2022low,gamgam2023satisfying}. \cite{al2023self} showed that RIS assignment can reduce the power consumption in RIS-aided NOMA-TDMA. The work in \cite{lei2023hybrid} demonstrated that RIS beamforming optimizing could improve the minimum beampattern gain (MBBG) in the hybrid NOMA-TDMA-aided integrated sensing and communication (ISAC) system. In \cite{lyu2023hybrid}, the combination of STAR-RIS and hybrid NOMA-TDMA outperforms only RIS-based and the OMA-based frameworks in terms of maximal of minimum downlink rate. \cite{chen2022active} showed the tradeoff between performance and cost can be more effectively balanced in the hybrid NOMA-TDMA with the assistance of active RIS.
In \cite{zhu2022low}, the benefits of NOMA and TDMA manifest the reduction of the total delay in the finite blocklength (FBL) regime. \cite{gamgam2023satisfying} showed that the timely throughput can be improved up to 46\% by the hybrid NOMA compared to conventional OMA. 
However, to the best of our knowledge, the study of optimizing RIS reflecting beamforming in the hybrid NOMA-TDMA-aided URLLC networks is still missing in the literature.

Therefore, in this paper, we consider combining the uplink NOMA scheme within the user pairs and the TDMA scheme between the user pairs. In particular, how to address the trade-off mentioned above between performance and complexity
by optimally allocating the transmit power, blocklength of users with limited radio resources, receiving beamforming, and passive beamforming is still an open problem. This issue is especially critical in multi-user scenarios, where the scalability of the allocation schemes should also be
considered. 
Additionally, because URLLC is mostly used to send measurement data or control signals with small packet sizes, throughput optimization is less important. Therefore, our goal is to generalize the min-max decoding-error probability-based resource allocation, which simultaneously reflects the minimization of the decoding-error probability for all devices and the fairness among them—a factor that has not yet been taken into account in previous research on RIS-URLLC. 
Our main contributions are summarized as follows:
\begin{itemize}
    \item  In order to minimize the maximum decoding error probability for all users, we consider joint power allocation, transmission blocklength, receiving beamforming, RIS reflection, and user pairing optimization. This optimization is conducted while taking into account limits on transmit power and latency. The formulated problem above poses significant challenges because of the non-convex nature and the complexity associated with throughput expression in the small blocklength regime. In order to achieve this objective, the formulated problem is divided into three separate subproblems.

   \item The first subproblem is decoding order determinations. The SDR was devised in order to determine the decoding orders by identifying the RIS reflection that maximizes the combined channel gain. Subsequently, utilizing the established decoding sequence, the whole user population is categorized into two distinct cohorts: proficient users and less proficient users. In the non-orthogonal multiple access (NOMA) system, data transmission to the base station (BS) occurs in each time slot through the random pairing of one strong and one weak user.
   \item The second subproblem involves the combined optimization of power allocation, receiving beamforming, RIS reflection, and transmission blocklength while maintaining fixed user pairing. In this study, we present a novel approach for addressing the subproblem in an efficient manner. Our proposed algorithm combines the methodologies of sequential convex approximation (SCA) and semidefinite relaxation (SDR) to achieve this goal.
   \item The third subproblem is determining user pairing for each time slot. In order to address the problem, we suggest employing a Hungarian matching-based iterative method.
   \item Finally, by the implementation of a comprehensive simulation, we are able to demonstrate the notable benefits of our suggested framework in comparison to alternative benchmark methods.
\end{itemize}
The rest of this paper is organized as follows. Our system model is presented in Section \ref{sec_sm}. Problem formulation is provided in Section \ref{sec_pf}, and Section \ref{sec_sr} represents our proposed solution. The numerical results are presented in Section V. The paper is finally concluded in Section \ref{sec_con}. \\

\textit{Notation}: 
Boldface lower-case and upper-case characters stand for vectors and matrices, respectively. For real-valued elements, the set of $n\times m$ matrices is represented by $\mathbb{R}^{n\times m}$, and for complex-valued entries, by $\mathbb{C}^{n\times m}$. $\mathbb{S}_+^n$ is a representation of the set of $n\times n$ semidefinite positive matrices. The diagonal matrix with a diagonal vector $\ba$ is represented by $\diag(\ba)$, the $(n,m)$th entry of a matrix $\bA$ is indicated by $[\bA]_{n,m}$, and the $n$th entry of a vector $\ba$ is shown by $[\ba]_n$.

\section {System Model}\label{sec_sm}
\begin{figure}[t]
\centering
    \includegraphics[width= \linewidth ]{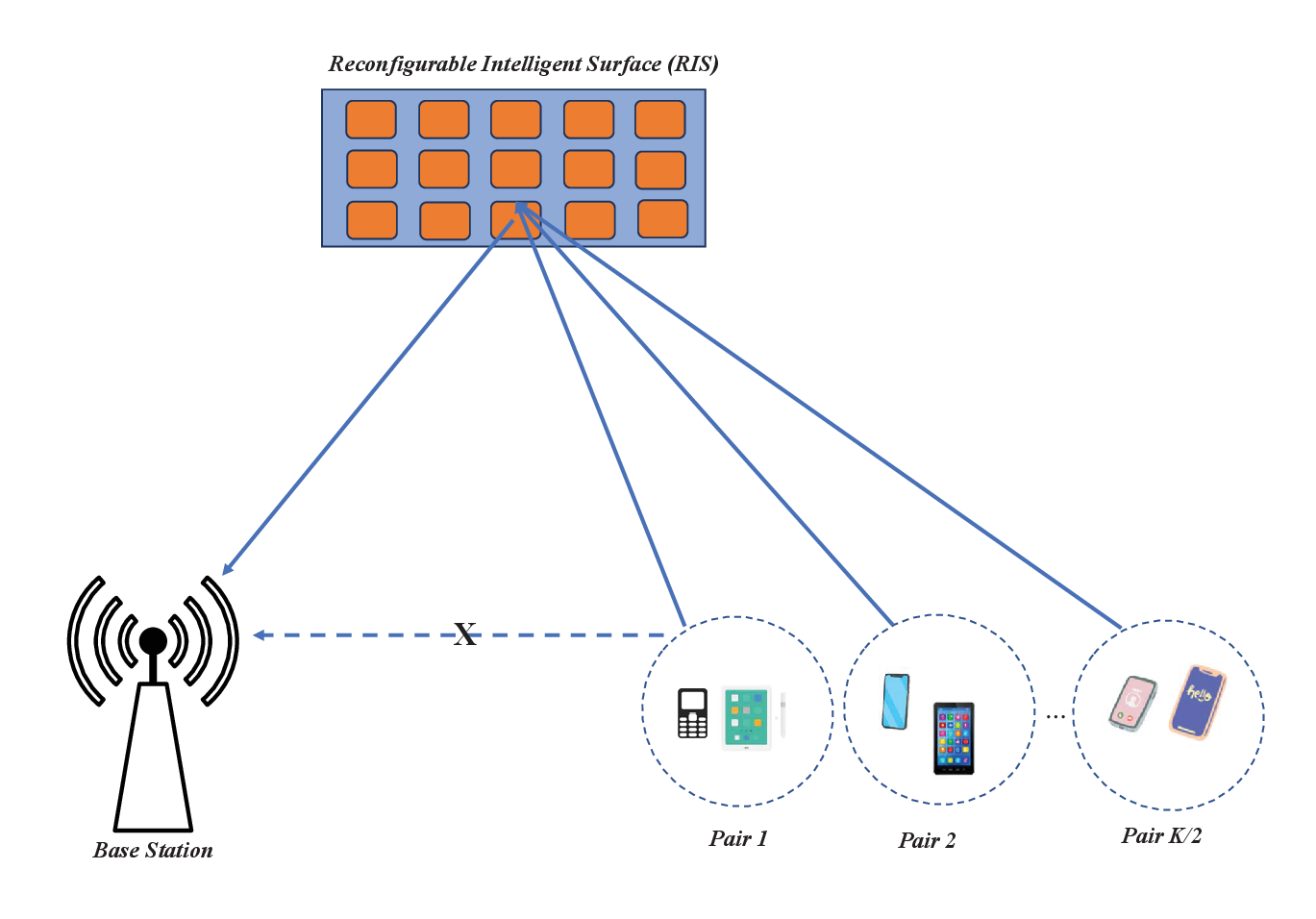}
	\caption{{System model.}}
	\label{fig_system}
\end{figure}

In this paper, we consider a RIS-URLLC, as shown in Fig. \ref{fig_system}, whereby several users operate covertly from the base station (BS) and employ short blocklengths for transmission, aided by a RIS. There exists a set $\mathcal{K}$ of $K$ users, each equipped with a single antenna, engaging in transmission to a BS that possesses $N_t$ antennas. 
 All users transmit status update packets, respectively with $D_k$-bits $(k \in \mathcal{K} = \{1, \cdots, K\})$ information to the BS. The direct connections between users and BS are impeded due to several constraints. In order to improve the communication link between users and BS, RIS consisting of $N$ components is employed. The $K$ users are partitioned into $K/2$ groups, with each group consisting of two individuals. Simultaneous transmission occurs among users within the same group, whereas distinct groups employ a TDMA approach for transmission. In our paper, we make the assumption that the channel gain remains constant during the specified time length.

User $k$ encodes the $D_k$ information into a single packet (codeword) with a length of $M^2$ symbols. The channels between BS and RIS, and between RIS  and user $k$  are denoted by  $\bH \in {\mathop{\mathbb{C}}}^{N\times N_t} $, and $\boldsymbol{f}_{k} \in {\mathop{\mathbb{C}}}^{N \times 1}$, respectively.
We can model $\bH$ as the following Rician fading channel
\begin{equation}
    \bH = \sqrt{\frac{\rho_0}{d_0^{\alpha_1}}}\left( \sqrt{\frac{K_1}{1+K_1}} \bH^{LOS} + \sqrt{\frac{1}{1+K_1}} \bH^{NLOS}\right),
\end{equation}
where the deterministic line-of-sight (LoS) component is denoted as $\bH^{LOS}$, while the random non-line-of-sight (NLoS) component is represented as $\bH^{NLOS}$. Both components are modeled using Rayleigh fading. $d_0$ represents the distance between BS and RIS.
$\alpha_1 \geq 2$ is used to express the path loss exponent, whereas
$\rho_0$ is used to describe the path loss at a reference distance of $1$ m. $K_1$ represents the Rician factor. Similarly, the Rician fading channel from the RIS to user $k$ is expressed as:

\begin{equation}
 \boldsymbol{f}_k = \sqrt{\frac{\rho_0}{d_k^{\alpha_2}}}\left( \sqrt{\frac{K_2}{1+K_2}} \boldsymbol{f}_k^{LOS} + \sqrt{\frac{1}{1+K_2}} \boldsymbol{f}_k^{NLOS}\right),   
\end{equation}
where $\boldsymbol{f}_k^{LOS}$ and $\boldsymbol{f}_k^{NLOS}$ are the LoS and NLoS components, respectively, and $d_k$ denotes the distance between the RIS and user $k$. $\alpha_2$ and $K_2$ are path loss exponent and Rician factor, respectively. The reflection vector $\boldsymbol{\phi} \in \mathbb{C}^{N\times 1}$ of RIS is determined by the phase shift vector $\bth=[\theta_{1},\theta_{2},\cdots,\theta_{N}]^T \in \mathbb{R}^{N\times 1}$ as 
\begin{equation}
	\boldsymbol{\phi}=[\phi_{1},\phi_{2},..., \phi_{N}]^T= [e^{j\theta_{1}},e^{j\theta_{2}},...,e^{j\theta_{N}}]^T,
\end{equation}
where $\theta_{n} \in [0,2\pi)$ leading to the unit modulus reflection, or equivalently $|\phi_{n}|=1$  and $n \in \mathcal{N} \triangleq \{1,2,\cdots,N\}$. The received signal at BS  can be written as
\begin{equation}
\label{eq_signal_u}
	z_k=\sum_{k \in \mathcal{K}}\bw_k^H\left(\left\{\bH^{H} \diag(\bphi) \boldsymbol{f}_{k}\right\}s_k+\bn_0\right),
\end{equation} 
where $s_k$ is the symbol delivering the model of user $k$, $\bn_0 \sim \mathcal{C}\mathcal{N}(0,\sigma^2)$ is the additive white Gaussian noise (AWGN) with the noise power $\sigma^2$ at BS, $\bw_k$ is receiving beamforming at BS of user $k$.
 We denote $k,k'$ are the same group if $b_{k,k'}=1$. Because each group has at most $2$ users, we have 
\begin{equation}
    \sum_{k \in \mathcal{K}|k'} b_{k',k} \leq 1,~\forall k' \in \mathcal{K}.
\end{equation}
The BS employs a technique known as successive interference cancellation (SIC) to accomplish symbol detection 
with \eqref{eq_signal_u}. The SIC order is denoted as $\pi(k)$ for user $k\in \mathcal{K}$, where $\pi(k)=n$ represents the detection of the symbol of user $k$ at the $n$-th order in the SIC process. Assuming that the decoding order $\pi$ corresponds to the decreasing order of direct channel gain, it can be inferred that the user signal with the greatest channel gain within each group is initially decoded at the BS. With $\pi(k) \leq \pi(k')$ and $k,k'$ are in same group, we have 
\begin{equation}
\label{eq_SINR_u}
	\gamma_{k}=\frac{
	\left|\bw_k^H\bH^{H}
 \diag(\bphi) \boldsymbol{f}_{k}\right|^2p_{k}}{\left|\bw_{k'}^H\boldsymbol{H}^{H} \diag(\bphi) \boldsymbol{f}_{k'}\right|^2 p_{k'} + \sigma^2},
\end{equation}
and 
\begin{equation}
\label{eq_SINR_u2}
	\gamma_{k'}=\frac{
	\left|\bw_{k'}^H\boldsymbol{H}^{H} \diag(\bphi) \boldsymbol{f}_{k'}\right|^2p_{k'}}{\sigma^2},
\end{equation}
where $p_k$ and $p_{k'}$ are the transmit power of user $k$ and $k'$, respectively.
Therefore, generally, the SINR of user $k$ is 
\begin{equation} \label{eq_sinr}
 \gamma_{k}= \frac{
	\left|\bw_k^H \bH^{H} \diag(\bphi) \boldsymbol{f}_{k}\right|^2p_{k}}{\sum\limits_{k' \in \mathcal{K}, \pi(k')>\pi(k)}\left|\bw_{k'}^H \boldsymbol{H}^{H} \diag(\bphi) \boldsymbol{f}_{k'}\right|^2 p_{k'} b_{k,k'}+ \sigma^2},   
\end{equation}
which leads to coding rate (in bits/sec/Hz) by user $k$ under short blocklength regime as follows \cite{polyanskiy2010channel}
\begin{equation}
    R_k=\log_2(1+\gamma_{k})-{\frac{V_k}{M}}\frac{Q^{-1}(\epsilon_k)}{\ln2}, 
\end{equation}
where $Q(\cdot)$ is the Q-function, $ V_k = (1-(1+\gamma_k)^{-2})^{1/2} \sim 1$ \cite{nasir2020min}. If a fixed number of $D_k$ bits are transmitted during the channel blocklength $M^2$ from user $k$, its coding rate becomes $R_k=\frac{D_k}{M^2}$ bits/sec/Hz.  Then the block error rate probability $\epsilon_k$ for user $k$ is given
\begin{equation}
    \epsilon_k=Q(g_k),
\end{equation}
where
\begin{equation}
 g_k=\frac{M\ln2}{{{V_k}}}\left(\log_2(1+\gamma_{k})-\frac{D_k}{M^2} \right), 
\end{equation}
The average energy of user $k$ is 
\begin{equation}
     \bar{E}_k =T_{sym}M^2p_{k},
\end{equation}
where $T_{sym}$ is the time for transmitting one symbol.

\section{Problem Formulation}\label{sec_pf}
This paper aims to formulate joint power allocation, receiving beamforming, blocklength, RIS reflection, and user pairing problem, which minimizes the maximal error decoding probability subject to the energy and the RIS’s practical reflection constraints. Thus, we write the formulated problem mathematically as follows:
\begin{subequations} \label{main_problem}
	\begin{align}
	\hspace{-1cm}	\textbf{P}: \quad      \underset{\boldsymbol{\phi},p_k,\bw_k, b_{k,k'},M}  {\min}&\max_k  \epsilon_k \tag{\ref{main_problem}}\\
		 \text{s.t.   } \quad  
		  &  0< p_{k} \leq p_{k}^{\max},  ~\forall k \in \mathcal{K}  \label{P1_cons1},\\
		  &   |\phi_{n}|=1,  ~\forall n \in \mathcal{N}, \label{P1_cons2} \\
            & ||\bw_k||^2=1, ~\forall k \in \mathcal{K}  \label{P1_cons7},\\
		  &  \sum_{k \in \mathcal{K}|k'} b_{k',k} \leq 1,~\forall k, k' \in \mathcal{K} \label{P1_cons3},\\	
		  & \sum_{k \in \mathcal{K}} \bar{E}_k \leq E_0 \label{P1_cons5},\\
		  & M^2 \leq \frac{T}{T_{symbol}}  \label{P1_cons6},\\
            &  b_{k',k} \in \{0,~1\}, ~\forall k, k' \in \mathcal{K} \label{P1_cons4},\\
            & M^2 \in \mathbb{Z}^+  \label{P1_cons8}, 
   \end{align}		
\end{subequations}
where $T$ is the duration for one-time slot, $p_{k}^{\max}$ is the maximum transmit power of user $k$. Constraints (\ref{P1_cons1}) and (\ref{P1_cons2}) are the power constraint for each user and the RIS reflection constraints of the RIS’s elements, respectively. Constraint (\ref{P1_cons7}) presents the receiving beamforming constraint, and constraint (\ref{P1_cons3}) ensures each user can be paired with at most one other user. Constraint (\ref{P1_cons5}) expresses the total energy constraint for all users, and constraint (\ref{P1_cons6}) is the latency constraint. Constraint (\ref{P1_cons4}) ensures the user pairing constraint. Finally, constraint (\ref{P1_cons8}) ensures that the blocklength allocated to each user is a positive integer. On the other hand, $\epsilon_k=Q(g_k)$ and Q-function is a monotonically decreasing function. 
Then, problem \textbf{P} can be replaced by the max-min function of  $g_k$.
\begin{subequations} \label{main_problema}
	\begin{align}
	\hspace{-1cm}	\textbf{P1}: \quad  & \underset{\boldsymbol{\phi},p_k,\bw_k, b_{k,k'},M} {\max} \min_k  ~ g_k \tag{\ref{main_problema}}\\
		 \text{s.t.   } \quad  
		  & \text{(\ref{P1_cons1}),~(\ref{P1_cons2}),~(\ref{P1_cons7}),~(\ref{P1_cons3}),~(\ref{P1_cons4}),~(\ref{P1_cons5}),~(\ref{P1_cons6}),~(\ref{P1_cons8})} \notag.
	\end{align}		
\end{subequations}
The formulated problem \textbf{P1} involves the continuous and integer variables. Furthermore, a strong coupling of the optimization variables and the intra-pair interference of a NOMA pair exists. As a consequence, \textbf{P1} is a non-convex MINLP problem that is challenging to solve. As a result, we develop an efficient solution to solve problem \textbf{P1}.

\section{Proposed Solution}\label{sec_sol}
 In this regard, we first relax the mixed integer problem \textbf{P1} by removing the integer constraints (\ref{P1_cons8}) on the variables
$M$. To facilitate the solution design and to smooth the objective
function of \textbf{P1}, an auxiliary variable $\chi$ is introduced so that
the equivalent form of \textbf{P1} 
\begin{subequations} \label{main_problemb}
	\begin{align}
	\hspace{-1cm}	\textbf{P2}: \quad  & \underset{\boldsymbol{\phi},p_k,\bw_k, b_{k,k'},M,\chi} {\max} \chi \tag{\ref{main_problemb}}\\
		 \text{s.t.   } \quad 
            &g_k \geq \chi,  ~\forall k \in \mathcal{K},\label{P2_cons1}\\
		  & \text{(\ref{P1_cons1}),~(\ref{P1_cons2}),~(\ref{P1_cons7}),~(\ref{P1_cons3}),~(\ref{P1_cons4}),~(\ref{P1_cons5}),~(\ref{P1_cons6}).} \notag
	\end{align}		
\end{subequations}
In order to implement the NOMA communication system, it is important to arrange the users in accordance with their respective effective channel gains.
However, In the case of RIS-assisted NOMA, it is important to note that the combined gain 
$\bH^{H} \diag(\bphi) \boldsymbol{f}_{k}$ s influenced by the reflection
coefficient $\bphi$. previous studies have demonstrated that NOMA systems employing fixed power distribution techniques exhibit superior performance when pairing two users with differing channel conditions. Therefore, we propose an approach to determine the decoding orders by finding RIS reflection vectors that maximize the sum of combined channel gains. Based on the objective of accomplishing decoding orders, the entire user population is categorized into two distinct groups: strong users and weak users. Subsequently, each pair is formed by pairing one strong user with one weak user. Furthermore, due to the inherent challenges associated with handling both continuous and integer variables in \textbf{P2} is rather difficult to tackle, we employ the bi-level optimization framework. This allows us to decompose \textbf{P2} into two sub-problems: the inner problem focuses on optimizing power allocation, receiving beamforming, blocklength, and RIS reflection, while the outer problem deals with user pairing. The inner problem corresponds to a max-min $g_k$ problem concerning the power allocation, receiving beamforming, blocklength, and RIS reflection variables given certain user pairing relationships. The
outer problem corresponds to max-min $g_k$ considering the user pairing variable. Consequently, the proposed algorithm has three steps, which are shown as follows. \\

\begin{itemize}
\item The first step is to determine the decoding order and then divide the total users into strong user subsets and weak user subsets.
\item The second step is to optimize power allocation, receiving beamforming, RIS reflection, and blocklength under a fixed user pairing variable.
\item The third step is to achieve the optimal user pairing relationships between a weak user and a strong user with the optimizing variables obtained by the second step.
\end{itemize}

\subsection{Decoding Order Determination}
This subsection presents a low-complexity algorithm to determine the decoding orders based on maximizing the combined channel gain. The formulated optimization problem is as follows:
\begin{subequations} \label{prob_or1}
	\begin{align}
	\hspace{-1cm}	\textbf{P3}: \quad  \underset{\bphi}\max & ~ |\bH^{H}
 \diag(\bphi) \boldsymbol{f}_{k}|^2 \tag{\ref{prob_or1}}\\
		 \text{s.t.   } \quad  
		  &  ~\text{(\ref{P1_cons2}).} \notag
	\end{align}		
\end{subequations}
In problem \textbf{P3}, the quadratic form $|\bH^{H}
 \diag(\bphi) \boldsymbol{f}_{k}|^2$
can be rewritten as $\bphi^H \bR_k \bphi$, where $\bR_k=\text{diag}((f_k)^H)\bH^H\bH\text{diag}(f_k)$.
By introducing slack matrices $\bS = \bphi \bphi^H$,
which are rank-one and positive semidefinite (PSD), we have $|\bH^{H}
 \diag(\bphi) \boldsymbol{f}_{k}|^2= \text{tr}(\bR_k \bS)$. Thus, problem \textbf{P3} can be transformed to
\begin{subequations}\label{problem_or}
    \begin{align}
	\hspace{-1cm}	\textbf{P3-1}: \quad  & \underset{\boldsymbol{S}} {\max} ~\text{tr}(\bR_k \bS) \tag{\ref{problem_or}}\\
		 \text{s.t.   } \quad 
            &\text{rank}(\boldsymbol{S})=1, \label{problem_or_1}\\
            & \boldsymbol{S} \succeq 0, \label{problem_or_2}.		  
	\end{align}	   
\end{subequations} 
As a standard SDP problem, problem \textbf{P3-1} can be solved by existing solvers in CVX \cite{cvx}. However, to deal with the challenge of the rank-one constraint (\ref{problem_or_1}), we can apply the Gaussian randomization to obtain a suboptimal solution \cite{Yang2021intelligent}. Based on the searched reflection matrix $\bphi^*$, if the combined channel gains experienced by any two users $(k,k')$ associated with BS can be arranged as $|\bH^{H} \diag(\bphi) \boldsymbol{f}_{k'}|^2 \leq  |\bH^{H} \diag(\bphi) \boldsymbol{f}_{k}|^2$. Then, we choose $K/2$ users with the most effective channel gain as the strong users and the remaining $K/2$ users as the weak users. In each time slot of TDMA, one strong user $k$ and one weak user $k'$ are paired and transmit data to BS in the NOMA scheme with the decoding order $\pi(k) \leq  \pi(k')$.
\subsection{Power Allocation, Receiving Beamforming, RIS Reflection, and Blocklength Optimization with Random User Paring}
In this section, we investigate Alternating Optimization (AO)
approach to handle problem \textbf{P1} efficiently by dividing it into
four subproblems. \\

\subsubsection{ Power Allocation Optimization with Fixed Receiving Beamforming, RIS Reflection, and Blocklength}
\begin{subequations} \label{main_problem3}
	\begin{align}
	\hspace{-1cm}	\textbf{P4}: \quad  \underset{\bp,\chi}\max & ~ \chi \tag{\ref{main_problem3}}\\
		 \text{s.t.   } \quad  
		  &  ~\text{(\ref{P2_cons1}), ~(\ref{P1_cons1}),~(\ref{P1_cons5}),} \notag
	\end{align}		
\end{subequations}
where $\bp=\{p_k, \forall k \in \mathcal{K}\}.$
The constraint (\ref{P2_cons1}) is equivalent to 
\begin{equation}\label{P2a_cons1}
   2^{\frac{\chi} {M\ln2} +\frac{D_k}{M^2}}-1  \leq \gamma_{k}.
\end{equation}
In addition, (\ref{eq_sinr}) 
is equivalent to 
\begin{equation}
\label{eq_SINR_u_app1}
	 {\sum\limits_{k'\in \mathcal{K}, \pi(k')>\pi(k)}}{X_{k'}^k p_{k'} \gamma_{k}+ \sigma^2}
	= { X_k p_{k}}, 
\end{equation}
where $X_k=\left|\bw_k^H \boldsymbol{H}^{H} \diag(\bphi) \boldsymbol{f}_{k}\right|^2$ and $X_{k'}^k=\left|\bw_{k'}^H \boldsymbol{H}^{H} \diag(\bphi) \boldsymbol{f}_{k'}\right|^2 $.
We replace $p_{k'} \gamma_{k}$ with its upper bound defined as \cite{beck2010sequential}
\begin{equation}
    p_{k'} \gamma_{k} \leq \frac{\lambda_{k,k'}}{2}\gamma_{k}^2+\frac{1}{2\lambda_{k,k'}} p_{k'}^2, 
\end{equation}
where $\boldsymbol{\lambda}=\{\lambda_{k,k'}| k,k' \in \mathcal{K}\}$. %
To make problem (\ref{main_problem3}) be convex, we replace (\ref{eq_SINR_u_app1}) by 
\begin{equation}\label{P2a_cons2}
     X_k p_{k} \geq {\sum\limits_{k'\in \mathcal{K}, \pi(k')>\pi(k)}} {X_{k'}^k (\frac{\lambda_{k,k'}}{2}\gamma_{k}^2+\frac{1}{2\lambda_{k,k'}} p_{k'}^2)+ \sigma^2}. 
\end{equation}
Then problem (\ref{main_problem3}) is equivalent to 
\begin{subequations} \label{main_problem2aa}
	\begin{align}
	\hspace{-1cm}	\textbf{P4-1}: \quad  \underset{\bp,\chi, \boldsymbol{\gamma}} \max & ~ \chi \tag{\ref{main_problem2aa}}\\
		 \text{s.t.   } \quad  
		  &  ~\text{(\ref{P2a_cons1}),~(\ref{P2a_cons2}), ~(\ref{P1_cons1}),~(\ref{P1_cons5}).} \notag
	\end{align}		
\end{subequations}
where $\boldsymbol{\gamma}=\{\gamma_k, \forall k \in \mathcal{K}\}$.

It can be observed that the objective function and all constraints in problem \textbf{P4-1} exhibit convexity. As a result, the Karush-Kuhn-Tucker (KKT) solution of problem \textbf{P4-1} may be updated iteratively until convergence by optimally addressing its convex approximation problem using CVX \cite{cvx}. The algorithm presented in Algorithm \ref{alg_power} summarizes the proposed power allocation algorithm, which incorporates an adjustable convergence accuracy denoted as $\epsilon$.
\begin{algorithm}[t]
  \caption{SCA-Based Algorithm for Power Allocation}\label{alg_power}
  \begin{algorithmic}[1]
  \STATE \textbf{Input:} Receiving beamforming matrix $\{\bw_k\}$, RIS reflection $\bphi$, and blocklength $M$
  \STATE Set maximum number of iterations $I^p_{\max}$ and the maximum error tolerance $\epsilon^p$
  \STATE Initialize $\boldsymbol{p}_s^{(0)}$
  and  $\chi_s^{(0)}$. Set $i=0$. 
  \REPEAT
        \STATE  $i \leftarrow i + 1$.
        \STATE Obtain $\chi_s^{(i)}$ and   $\boldsymbol{p}_s^{(i)}$ by solving problem \textbf{P4-1}. 
  \UNTIL{($|\chi_s^{(i)}-\chi_s^{(i-1)}|<=\epsilon^p$ ) or ($i\geq I^p_{\max}$)}
\STATE \textbf{Output:} $\bp^o=\bp_s^{(i)}$ and $\chi^o=\chi_s^{(i)}$
 \end{algorithmic}
\end{algorithm}

\subsubsection{Receiving Beamforming Matrix Optimization}
For given power allocation, blocklength, and RIS reflection, the beamforming vectors $\bw=\{\bw_k, \forall k \in \mathcal{K}\}$ can be optimized by solving

\begin{subequations}\label{beamforming_prob}
    \begin{align}
    \hspace{-1cm}	\textbf{P5}: \quad  & \underset{\bw, \chi } {\max}    ~\chi \tag{\ref{beamforming_prob}}\\
    \text{s.t.   } \quad  
		  & \text{(\ref{P1_cons7}), (\ref{P2_cons1}).}\notag
    \end{align}
\end{subequations}
As (\ref{beamforming_prob}) is non-convex due to the non-convex constraint (\ref{P2_cons1}), we adopt the SDR technique to obtain an efficient approximate solution.  Denote the combined channel $\bq_k =
\boldsymbol{H}^{H} \diag(\bphi) \boldsymbol{f}_{k}$. By introducing $\bQ_k = \bq_k \bq_k^H$ and $\bW_k = \bw_k \bw_k^H$ , the term $|\bw_k^H\boldsymbol{H}^{H} \diag(\bphi) \boldsymbol{f}_{k}|^2$ can be written as $\text{Tr} (\bW_k \bQ_k)$, where $\bW_k$ needs to satisfy $\bW_k \succeq 0$ and $\text{rank}(\bW_k) = 1$. Since the constraint $\text{rank} (\bW_k) = 1$ is nonconvex, we relax this constraint and (\ref{beamforming_prob}) can be transformed as the following problem
\begin{subequations}\label{beamforming_prob2}
    \begin{align}
    \hspace{-1cm}	\textbf{P5-1}: \quad   &\underset{\bW, \chi } {\max}    ~\chi \tag{\ref{beamforming_prob2}}\\
    \text{s.t.} \quad & \eta_k  \bigg( \!\!\!\! \sum_{\pi(k') > \pi(k)}\!\!\!\!\! \text{Tr}(\bW_{k'} \bQ_{k'})p_{k'} +\sigma^2 \bigg)	   \notag \\
     & \quad \leq \text{Tr}(\bW_k \bQ_{k})p_{k}, \forall k \in \mathcal{K}, 	\label{cons1_pb2}\\
        &  \bW_k \succeq 0, \forall k \in \mathcal{K}, \label{cons2_pb2}\\
		&\text{Tr}(\bW_k)=1, \forall k \in \mathcal{K} \label{cons3_pb2},
     \end{align}		
\end{subequations}
where $\bW=\{\bW_k, \forall k \in \mathcal{K} \}$.
\begin{algorithm}[t]
    \caption{Bisection Search and {SDR}-Based Algorithm for Receiving Beamforming Matrix Optimization}
    \label{alg_beamforming}
    \begin{algorithmic}[1]
    \STATE \textbf{Input:} Transmission power $\boldsymbol{p}$, RIS reflection $\bphi$, and blocklength $M$
    \STATE Set maximum error tolerance $\epsilon^b$
    \STATE Initialize $\chi_{\min}$ and $\chi_{\max}$.  \             
        \WHILE{$\chi_{\max}-\chi_{\min} \geq \epsilon^b $}
			\STATE Solve problem (\ref{beamforming_prob2}) with $\bar{\chi}=\frac{\chi_{\min}+\chi_{\max}}{2}$.
			 \IF{ problem (\ref{beamforming_prob2}) is feasible}
			  \STATE $\boldsymbol{W}^\dagger \leftarrow \boldsymbol{W}_{\bar{\chi}}$.
			  \STATE $\chi_{\min} \leftarrow \bar{\chi}$.
			  \ELSE 
			  \STATE  $\chi_{\max} \leftarrow \bar{\chi}$.
			 \ENDIF
		\ENDWHILE 
      \FOR{$k= 1, 2, \cdots,K$}
		  \FOR{$c=1,2,\cdots,C$}
		  \STATE Generate a Gaussian random vector $\tilde{\bw_k}^{(c)} \sim \mathcal{CN}(\boldsymbol{0},\bW_k^\dagger)$. 
		  \STATE Obtain the objective value $\chi^{(c)}$ of (\ref{beamforming_prob2}) with $\bw_k=\bw_k^{(c)}$. 
		  \ENDFOR
	      \STATE Find the best candidate  $c^\dagger=\arg\max_{1\leq c \leq C} \chi^{(c)}$.
            \STATE $\bw_k^\dagger=\bw_k^{(c^\dagger)}$
       \ENDFOR
	      	\STATE \textbf{Output:} $\bw^\dagger=\{\bw_k^{(\dagger)}, \forall k \in \mathcal{K}\}$       
    \end{algorithmic}
\end{algorithm}
However, (\ref{beamforming_prob2}) is still non-convex due to the non-convex constraint (\ref{cons1_pb2}) and (\ref{cons2_pb2}). We first drop the rank-one constraint (\ref{cons2_pb2}). The bisection-search method is used to solve the coupling problem between $\chi$ and $\bW$, and then the SDR technique is used to obtain the optimal $\bW$. Specifically, with certain $\chi_{max}$ and $\chi_{min}$, we replace $\chi$ in $\eta_k$ in (\ref{beamforming_prob2}) by $\frac{\chi_{max}+\chi_{min}}{2}$ , and solve the resulting feasibility problem reduced from (\ref{beamforming_prob2}) as 

\begin{subequations}\label{beamforming_prob3}
	\begin{align}
	\hspace{-1cm}	\textbf{P5-2}: \quad  &\text{find} ~\bW=\{\bW_k, \forall k \in \mathcal{K}| \bW_k \in \mathbb{S}_+^{N_t}\}   \tag{\ref{beamforming_prob3}} \\
		\text{s.t.} \quad & \text{(\ref{cons1_pb2}),~(\ref{cons2_pb2}),~(\ref{cons3_pb2}).}
	\end{align}		
\end{subequations} 
The determination of the updated values for $\chi_{max}$ and $\chi_{min}$ is contingent upon the identification of a feasible solution $\bW$. It can be verified that when $\chi_{max}$ is sufficiently big and $\chi_{min}$ is sufficiently small, the bisection search described above for $\chi$ has the capability to produce a globally optimum matrix $\bW$ that is connected to beamforming in the $i$-th alternative iteration.
Next, it is necessary to contemplate the process of transforming the optimal solution $\bW^{*}$ into a viable solution $\{\hat{\bw_k}\}$ for the problem (\ref{beamforming_prob3}). The eigenvalue decomposition approach may be used to derive the optimum beamforming vectors $\{\bw_k^{*}\}$ when the optimal solutions $\bW^{*}$ of problem (\ref{beamforming_prob3}) adhere to the rank-one constraint. The Gaussian randomization approach is employed to acquire the approximate solutions of equation (\ref{beamforming_prob3}). 
To begin, we build a random vector $\hat{\bw_k}$ that conforms to a complex normal distribution with mean $0$ and variance $\bW_k^{*}$. Next, we utilize the operator $\hat{\bw_k}$ to address problem (\ref{beamforming_prob3}) and assess its viability. In this context, it is necessary to produce an ample quantity of feasible beamforming vectors $\hat{\bw_k}$, in an independent manner. Subsequently, the most optimal vector, denoted as $\hat{\bw_k}^{*}$, is selected from the pool of randomly generated vectors. Based on the above information, the final phase shift vector can be approximated as $\bw_k^{*} =\hat{\bw_k}^{*}$. The  Algorithm \ref{alg_beamforming} provides a step-by-step procedure for solving equation (\ref{beamforming_prob}) using an iterative approach.
\subsubsection{RIS Reflection Optimization}
For given blockength, receiving beamforming, and power allocation, the reflected vectors $\bphi$
can be optimized by solving
\begin{subequations}\label{phase_prob}
    \begin{align}
    \hspace{-1cm}	\textbf{P6}: \quad  & \text{find}    ~\bphi \tag{\ref{phase_prob}}\\
    \text{s.t.   } \quad  
		  & \text{(\ref{P1_cons2}), (\ref{P2_cons1}).}\notag
    \end{align}
\end{subequations}
The SDR technique, which is comparable to the approach outlined in Section IV-B-2, can be used to solve problem (\ref{phase_prob}). However, it takes a lot of time to solve the SDR problem, particularly when the matrix size is big. Therefore, we create an effective method based on SCA as follows in order to lower the computational complexity and ensure performance.
Denote $\bl_k=\bw_k^H \bH^H \text{diag}(f_k)$ and $\bL_k=\bl_k \bl_k^H$. 
By defining 
$\boldsymbol{S}={\boldsymbol{\phi}} {\boldsymbol{\phi}}^H \in\mathbb{S}_+^{N}$, 
With $\eta_k=2^{\frac{\chi} {M\ln2} +\frac{D_k}{M^2}}-1$, 
Problem (P2) can be written as
\begin{subequations}\label{phase_problem2}
    \begin{align}
	\hspace{-1cm}	&\textbf{P6-1}: \quad   \text{find}  ~\boldsymbol{S}  \tag{\ref{phase_problem2}}\\
		 \text{s.t.   } \quad 
            & \eta_k \bigg( \!\!\!\! \sum_{\pi(k') > \pi(k)}\!\!\!\!\! \text{tr}(L_{k'}\boldsymbol{S} )p_{k'} +\sigma^2 \bigg)	\leq \text{tr}(L_{k} \boldsymbol{S} )p_{k}, \forall k \in \mathcal{K}, \label{cons1_pp2}\\
            &\text{rank}(\boldsymbol{S})=1, \label{cons2_pp2}\\
            &[\boldsymbol{S}]_{n,n}	=1, ~ n=1,2,\cdots, N, \label{cons3_pp2}\\
            & \boldsymbol{S} \succeq 0 \label{cons4_pp2}.		  
	\end{align}	   
\end{subequations}
The equivalent form of the rank-one constraint (\ref{cons2_pp2}) can be expressed as
\begin{equation}\label{pp2}
    ||\bS||_*-||\bS||_2 \leq 0, 
\end{equation}
where $||\bS||_* = \sum_i \sigma_i (\bS)$ and $||\bS||_2 =\sigma_1 (\bS)$ denote the
nuclear norm and spectral norm, respectively, and $\sigma_i (\bS)$ is the
$i$th largest singular value of matrix $\bS$. For any $\bS \in  \mathbb{S}_+^{N}$,
we have $||\bS||_*-||\bS||_2 \leq 0$ and the equality holds if and
only if $\bS$ is a rank-one matrix. However, (\ref{phase_problem2}) is still a nonconvex constraint.
In order to take use of the penalty-based approach, we incorporate this restriction into the objective function of equation (\ref{phase_problem2}), which results in the optimization problem that follows:
\begin{subequations}\label{phase_problem3}
    \begin{align}
	\hspace{-1cm}	\textbf{P6-2}: \quad  &  \min_{\boldsymbol{S}}~ \frac{1}{2\mu}( ||\bS||_*-||\bS||_2) \tag{\ref{phase_problem3}}\\
		 \text{s.t.   } \quad & \text{(\ref{cons1_pp2}),~(\ref{cons3_pp2}),~(\ref{cons4_pp2})} \notag.           	  
	\end{align}	   
\end{subequations}
where $\mu$ is a penalty factor for (\ref{pp2}). As the result, using a modest $\mu$ will yield the minimum value of (\ref{pp2}).  It is important to remember that even if the rank-one constraint is relaxed in (\ref{phase_problem3}), the answer obtained by solving (\ref{phase_problem3}) is
an the best answer for $\mu \rightarrow 0$. Conversely, if $\mu$ is small enough, the solution to (\ref{phase_problem3}) is a rank-one solution. However, because of the objective function's DC form, (\ref{phase_problem3}) is not yet convex. We specify a lower bound of $||\bS||_2$, in order to address the DC form, and this may be done by
\begin{equation}
\begin{aligned}
    &||\bS||_2 \geq \bar{\bS}^{(n)}= || \bS^{(n)}||_2 \\
    &+ \text{tr} [u_{max} (\bS^{(n)}) (u_{max} (\bS^{(n)}))^H (\bS - \bS^{(n)})] ,  
\end{aligned}  
\end{equation}
where $u_{max} (\bS^{(n)})$ denotes the eigenvector corresponding to
the largest eigenvalue of $\bS^{(n)}$.
Accordingly, the optimization problem can be written as
follows:
\begin{subequations}\label{phase_problem4}
    \begin{align}
	\hspace{-1cm}	\textbf{P6-3}: \quad  &  \min_{\boldsymbol{S}}~ \frac{1}{2\mu}( ||\bS||_*-|| \bar{\bS}^{(n)}||)  \tag{\ref{phase_problem4}}\\
		 \text{s.t.   } \quad & (\text{\ref{cons1_pp2}}),~(\text{\ref{cons3_pp2}}),~(\text{\ref{cons4_pp2}}) \notag.           	 
    \end{align}	   
\end{subequations}
It is noted that the problem in (\ref{phase_problem4}) is convex, meaning optimization solvers like CVX may be used to solve it well. Algorithm {\ref{alg_phase_sca} provides a summary of the suggested penalty-based strategy for resolving the issue in (\ref{phase_problem4}).
\begin{algorithm}[t]
    \caption{{SCA}-Based Algorithm for RIS Reflection Optimization}
    \label{alg_phase_sca}
    \begin{algorithmic}[1]
    \STATE \textbf{Input:} transmission power $\boldsymbol{p}$, beamforming matrix $\{\bw_k\}$ and blocklength $M$ 
    \STATE Set maximum number of iterations $I^r_{\max}$ and the maximum error tolerance $\epsilon^r$
        \STATE Initialize $\bS^{(0)}$. Set $i=0$ 
        \REPEAT
        \STATE  $i \leftarrow i + 1$.
        \STATE Obtain $\bS^{(i)}$ by solving problem (\ref{phase_problem4}),
        \UNTIL{convergence or ($i\geq I^r_{\max}$)}
    \STATE  Decompose $\bS^{(i)} =\bphi^{(i)}(\bphi^{(i)})^H$         
   \STATE \textbf{Output:} $\bphi^o=\bphi^{(i)}$.       
    \end{algorithmic}
\end{algorithm}

\subsubsection{Blocklength Optimization}
For a given power allocation $\bp$, the beamforming vectors $\bw$ and reflect vector $\bphi$, we have the problem:
\begin{subequations} \label{block_problem}
	\begin{align}
	\hspace{-1cm}	\textbf{P7}: \quad  \underset{\chi, M}{\max} & ~ \chi \tag{\ref{block_problem}}\\
		 \text{s.t.   } \quad  & \text{(\ref{P1_cons5}),~(\ref{P1_cons6}),~(\ref{P2_cons1}).} \notag
	\end{align}		
\end{subequations}
It is observed that the problem in {(\ref{block_problem})} is a convex optimization problem that the CVX optimization toolbox can efficiently solve.
Since the integer constraint (\ref{P1_cons8}) was relaxed in problem \textbf{P1}, we adopt a greedy search method to solve the integer conversion problem \cite{nasir2020min}. Firstly, we initialize the integer solution as a  floor function of $M^2$. Next, we keep on allocating one blocklength to the user as long as the constraints (\ref{P1_cons5}), (\ref{P1_cons6}), (\ref{P2_cons1}) are met.
\subsection{User Pairing}
Existing research proved that the NOMA systems with
fixed power allocation schemes could perform better when pairing two users in distinct channel conditions. Then we consider the user pairing where one group has one strong user and one weak user. 
With the given power allocation, beamforming, phase shift, and blocklength optimization, problem \textbf{P1} can be transformed to
\begin{subequations} \label{pairing}
    \begin{align}
    \hspace{-1cm}	\textbf{P8}: \quad   \underset{\boldsymbol{b}} {\max}  &  ~\chi \tag{\ref{pairing}} \\
    \text{s.t.} \quad  
		  & \text{(\ref{P1_cons3}), ~(\ref{P1_cons4}),~(\ref{P2_cons1})}. \notag
    \end{align}
\end{subequations}
The transformed problem can be solved by the iterative bisection method to maximize the variable $\chi$. In each iteration, $i$, with the given $\chi_i$, the pairing scheme can be viewed as a matching problem in
a bipartite graph $\Omega_i (U_s, U_w, E)$, where $U_s, U_v$ are the strong and weak users, respectively and $E$ is the edges with vertexes from $U_s$ and $U_v$. The weight $e_{k,k'}$ of the edge between user $k \in U_s$ and user $k' \in U_w$ is based on the value of $g_k$ and $g_{k'}$ if $k$ and $k'$ are in the same group. Specially, $e_{k,k'}$ is equal to  $\min\{g_k,g_{k'}\}$ if $\min\{g_k,g_{k'}\} \geq \chi_i$ and 0, otherwise. Then, the maximum match of the bipartite graph $\Omega_i (U_s, U_w, E)$ with the given $\chi_i$ threshold is determined by employing the Hungarian algorithm\cite{kuhn2005hungarian}. 
The process is iterative until the difference between the upper bound $\chi_U$ and lower bound $\chi_L$ falls below the iteration accuracy $\epsilon^u$ or the maximum match turns into a perfect match, or the predefined maximum number of iterations is reached. Based on the matching when the iteration is converged, the NOMA groups are obtained, each of which is a pair of a strong user and a weak user. The Hungarian matching-based user pairing optimization is summarized in Algorithm \ref{alg_paring}.
\begin{algorithm}[t]
    \caption{Bisection and Hungarian Matching-Based User Pairing Algorithm}
    \label{alg_paring} 
    \begin{algorithmic}[1]
    \STATE \textbf{Input:} transmission power $\boldsymbol{p}$, beamforming matrix $\{\bw_k\}$, RIS reflection $\bphi$, blocklength $M$, and the corresponding $\chi^o$.
    \STATE Set maximum number of iterations $I^u_{\max}$ and the maximum error tolerance $\epsilon^u$
	\STATE Compute the weight matrix $\boldsymbol{\Omega}$ where each element is $e_{k,k'}=\min\{g_k,g_{k'}\}$ between the strong user $k$ and weak user $k'$.\\
        \STATE Initialize $\chi_H$ and $\chi_L$.
        \STATE Set $i=1$\\
               
        \WHILE {$\chi_H-\chi_L \geq \epsilon^u$ or $i \leq I^u_{\max}$ }
        \STATE Set $\chi_i=\frac{\chi_H+\chi_L}{2}$\\
        \STATE Build the edge $V$ between a strong user $k$ and a weak user $k'$ if $e_{k,k'} \geq \chi_i$ , generate the bipartite graph $\Omega$\\
        \STATE Obtain the maximum match $\Omega^i_{max}$ of $\Omega_{i}(U_s,~U_w,~E)$ by 
employing the Hungarian algorithm\\
              \IF{ $\Omega^i_{max}$ is not a perfect match}
			  \STATE $\chi_H=\chi_i$.\\
			   \ELSE 
			  \STATE  $\chi_L=\chi_i$.\\
			 \ENDIF
        \STATE $i\leftarrow i+1$\\
		\ENDWHILE  
   \STATE \textbf{Output:} $\Omega_{max}^o=\Omega^i_{max}$.

    \end{algorithmic}
\end{algorithm} \\

\subsection{Overall Algorithm, Convergence and Complexity Analysis}
\subsubsection{Overall Algorithm}
\begin{algorithm}[t]
    \caption{Proposed Three-Step Algorithm for Minimizing Maximal Decoding Error Probability }
    \label{alg_alternating} 
    \begin{algorithmic}[1]
    \STATE - \textbf{Step 1:  Design of decoding order and random user pairing}
    \STATE Solve problem (\ref{problem_or}) to obtain the combined channel gain of each user $|\bH^{H}
 \diag(\bphi) \boldsymbol{f}_{k}|^2$ .
    \STATE Obtain $\boldsymbol{\pi}=\{\pi(k), \forall k \in \mathcal{K}\}$ according to derived $|\bH^{H}
 \diag(\bphi) \boldsymbol{f}_{k}|^2$ and $K/2$ users with the most effective channel gain as the strong users and the remaining $K/2$ users as the weak users.
    \STATE Randomly divide $K$ users into $K/2$ groups, each has one strong user and one weak user.
    \STATE - \textbf{Step 2: Joint power allocation, receiving beamforming, RIS reflection, and blocklength optimization} 
       \STATE  Set maximum number of iterations $I_{max}$, and iteration index $i = 0$.
       \STATE Initialize $\bp^{(0)}$, $\bw^{(0)}$, $\bphi^{(0)}$ and $M^{(0)}$.
	\REPEAT
        \STATE $i \leftarrow i + 1$
		\STATE Given $M^{(i-1)}$, $\bw^{(i-1)}$ and $\bphi^{(i-1)}$, solve problem (\ref{main_problem2aa}) by applying
Algorithm \ref{alg_power}, and obtain $\bp^{(i)}=\bp^o$ and $\chi^o$.\\ 
		\STATE Given $\bp^{(i)}$, $M^{(i-1)}$ and $\bphi^{(i-1)}$, solve problem (\ref{beamforming_prob2}) by applying
Algorithm \ref{alg_beamforming} with $\chi_{\min}=\chi^o$, and obtain $\bw^{(i)}$.\\ 
            \STATE Given $\bp^{(i)}$, $M^{(i-1)}$ and $\bw^{(i)}$, solve problem (\ref{phase_problem4}) by applying
Algorithm \ref{alg_phase_sca}, and obtain $\bphi^{(i)}$.\\ 
            \STATE Given $\bp^{(i)}$, $\bw^{(i)}$ and $\bphi^{(i)}$, solve problem (\ref{block_problem}) by using CVX, and obtain $M^{(i)}$.\\ 
        \UNTIL {convergence or ($i\geq I_{max}$)}
	    \STATE - \textbf{Step 3: User pairing optimization}
	    \STATE Given $\bp^{(0)}$, $\bw^{(0)}$, $\bphi^{(0)}$ and $M^{(0)}$, solve problem (\ref{block_problem}) by Algorithm \ref{alg_paring}, and obtain $\boldsymbol{b}$.
    \end{algorithmic}
\end{algorithm}
The overall algorithm is summarized in Algorithm \ref{alg_alternating}, which consists of three steps: design of decoding order and random grouping, resource allocation, 
\subsubsection{Convergence}
{\color{black} Firstly, we prove the convergence of step 2. 
For given decoding order $\boldsymbol{\pi}$ and random grouping, we have 
\begin{equation}
 	\begin{aligned}
  	&	\chi(\bp^{(i-1)}, \bw^{(i-1)}, \bphi^{(i-1)}, M^{(i-1)}) \\
    & \overset{(a)}{\leq}   \chi(\bp^{(i)}, \bw^{(i-1)}, \bphi^{(i-1)}, M^{(i-1)})   \\
      & \overset{(b)}{\leq}   \chi(\bp^{(i)}, \bw^{(i)}, \bphi^{(i-1)}, M^{(i-1)}) \\
      & \overset{(c)}{\leq}  \chi(\bp^{(i)}, \bw^{(i)}, \bphi^{(i)}, M^{(i-1)})\\
      & \overset{(d)}{\leq}  \chi(\bp^{(i)}, \bw^{(i)}, \bphi^{(i)}, M^{(i)}).       
 		\end{aligned}	
  \end{equation}
where $\chi(\bp^{(i-1)}, \bw^{(i-1)}, \bphi^{(i-1)}, M^{(i-1)}) $ is the value of the objective function of problem (\ref{main_problemb}) at iteration $(i-1)$ and (a), (b), (c), (d) results from Algorithm \ref{alg_power}, \ref{alg_beamforming}, \ref{alg_phase_sca} and \ref{alg_paring}, respectively. As a result, the objective function is non-decreasing after each iteration. Moreover, since the objective is continuous over the compact feasible set of problem \textbf{P1}, it is upper-bounded by some finite positive number \cite{boyd2004convex}. Hence, Step 2 is ensured to converge. Step 3 involves both bisection and Hungarian, and then it converges.
 }
\subsubsection{Complexity Analysis}
{\color{black} The computational complexity of Algorithm \ref{alg_alternating} is
analyzed as follows. 
Step 1 deals with solving the semidefinite program
(SDP) problem, then, the complexity of Algorithm (\ref{main_problemb}) is $\mathcal{O}(M^{4.5})$ \cite{luo2010semidefinite}. In the step 2, the complexities of Algorithm \ref{alg_power}, \ref{alg_beamforming}, \ref{alg_phase_sca} are 
$\mathcal{O} (I^p_{\max}(K+1)^3)$, $\mathcal{O}((M+1)^{4.5}\log(1/\epsilon^b))$, $\mathcal{O}(I^r_{max}(M+1)^{4.5})$, respectively. The complexity of obtaining optimal blocklength is $\mathcal{O}(I_{M})$, where $I_M$ is the number of iterations needed for convergence. Thus, the complexity of step 2 is $\mathcal{O}(I_{\max}( I^p_{\max}(K+1)^3+ (M+1)^{4.5}\log(1/\epsilon^b)+ \mathcal{O}(I^r_{max}(M+1)^{4.5}) ))$. he computational complexity of the iterative bisection approach in step 3 is $\mathcal{O}(\log(1/\epsilon^u))$ while the Hungarian method's computational complexity in each iteration is $\mathcal{O}(\frac{K^3}{8})$ \cite{kim2014iterative}. 
 Thus, the computational complexity of Algorithm \ref{alg_paring} in step 3 is $\mathcal{O}(\frac{K^3}{8}\log(1/\epsilon^u))$. Totally, the complexity of Algorithm \ref{alg_alternating} is $\mathcal{O}(M^{4.5}+I_{\max}( I^p_{\max}(K+1)^3+ (M+1)^{4.5}\log(1/\epsilon^b)+ \mathcal{O}(I^r_{max}(M+1)^{4.5}) )+\frac{K^3}{8}\log(1/\epsilon^u))$.
 }

\section{Simulation Results}\label{sec_sr}
This section presents numerical findings that illustrate how well the suggested NOMA-TDMA transmission system performs while working with RIS to reduce the maximum error coding error. We examine a coordinate system that is three-dimensional and has the axes $(x, y, z)$. We assume that the RIS and BS have elevations of $10$ m, and that their horizontal coordinates are $(0, 0)$ and $(5, 15)$. The users are uniformly and randomly dispersed across the rectangular region created by the lines $x = 90, x = 190$, $y = - 10$, and $y = 10$. Their elevations are set to $0$ m. In every simulation, user coordinates are first generated at random and then fixed. 
In addition, we assume $\sigma^2 =-110$ dBm for the noise power at the BS and $p_{max} = 0.3$ W for the maximum transmit power of the user. The channel models include path loss and Rayleigh fading. The path loss exponents for the BS-RIS link and RIS-user are set to $\alpha_1 = 2.2, \alpha_2= 2.6$, and
the corresponding Rician factors are $K_1 = K_2 = 5$ dB. The maximum symbols of each user in the one-time slot $\frac{T}{T_{symbol}}$ is 20.

In Fig. \ref{fig_element}, a comparison is made between the worst-case error probability of the proposed scheme and many alternative baselines. These baselines include random phase, pure NOMA, NOMA without Hungarian pairing, location-based successive interference cancellation (SIC), and equal power. The comparison is conducted under the conditions where $K = 4, N_t = 3$, and $E_0 = 10$ W. In the context of pure NOMA, it is assumed that all four users are belonging to a single group, which simultaneously broadcasts their signals to the base station (BS). In the location-based SIC system, users are categorized into two categories according to the proximity of their locations to the base station (BS). Furthermore, power is distributed evenly in the equal power scheme. In contrast, in the Hungarian pairing scheme, power allocation, receiving beamforming, RIS reflection, and blocklength optimization are concurrently optimized, but users are paired randomly. The initial observation reveals that the worst-case decoding error in all schemes exhibits a decrease as the parameter $N$ increases. This may be attributed to the fact that a bigger size of reconfigurable intelligent surface (RIS) can leverage the potential advantage provided by a higher reflecting array. Furthermore, the proposed scheme has the lowest minimum-maximum error when compared to the six designs that have been taken into consideration. This is due to its ability to enhance power optimization, receiving beamforming, RIS reflection, and user pairing. Furthermore, the proposed scheme demonstrates a reduction in the worst-case decoding error when compared to the pure NOMA scheme. This improvement may be attributed to the fact that the proposed method encounters a lower level of inter-group interference in comparison to the pure NOMA scheme. The results indicate that the NOMA-TDMA scheme has superior performance compared to pure NOMA in terms of minimizing the maximal decoding error.
\begin{figure}[t]
\centering
    \includegraphics[width=0.98\linewidth ]{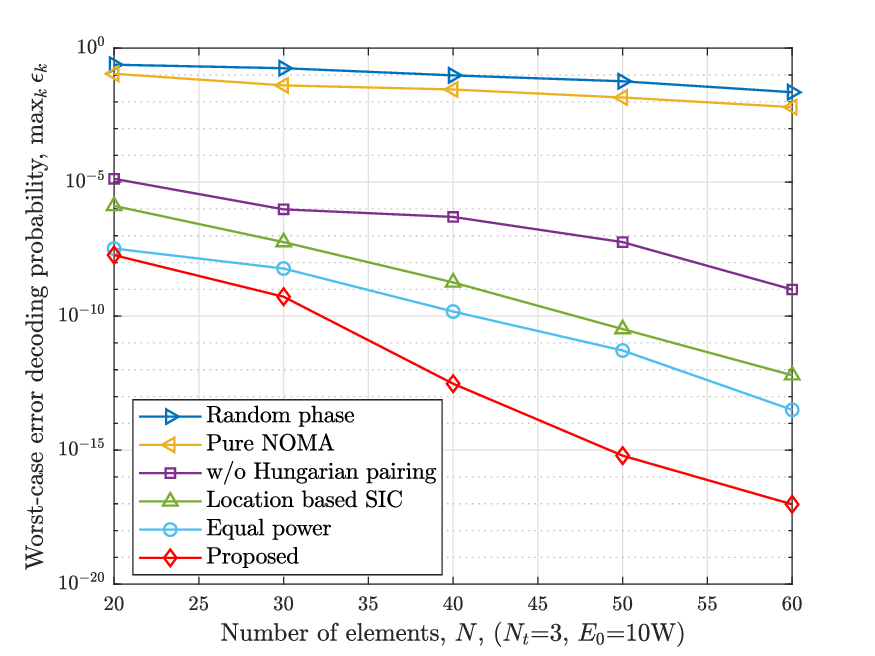}
	\caption{{\color{black} Optimized min-max decoding-error probability under different schemes.}}
	\label{fig_element}
\end{figure}
\begin{figure}[t]
\centering
    \includegraphics[width=0.98\linewidth ]{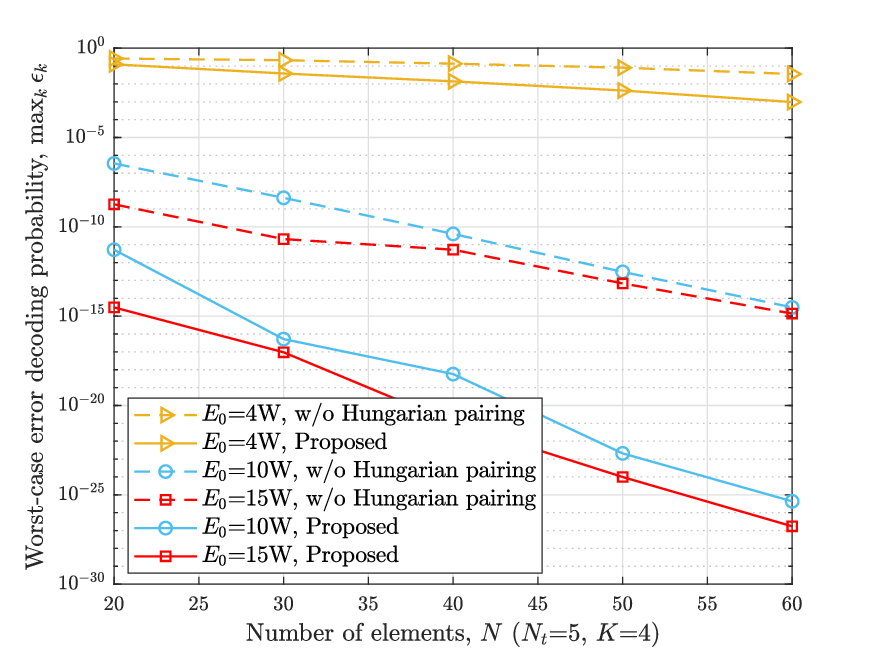}
	\caption{{\color{black} Optimized min-max decoding-error probability versus
the number of elements of the RIS when $E_0$ varies.}}
	\label{fig_NvsNoma}
\end{figure}

Fig. \ref{fig_NvsNoma} shows the worst-case error decoding error when the number of element $N$ varies under different total energy constraints $E_0$. It is observed that the decoding error in the worst-case scenario lowers as the values of $N$ and $E_0$ grow. The increase in power allocation to users as E0 grows leads to an increase in the signal-to-interference-plus-noise ratio (SINR) of users in the worst-case scenario. Consequently, the performance of the min-max decoding algorithm deteriorates. The efficiency of the suggested Hungarian matching-based user pairing is demonstrated in comparison to random pairing.
\begin{figure}[t]
\centering
    \includegraphics[width=0.98\linewidth ]{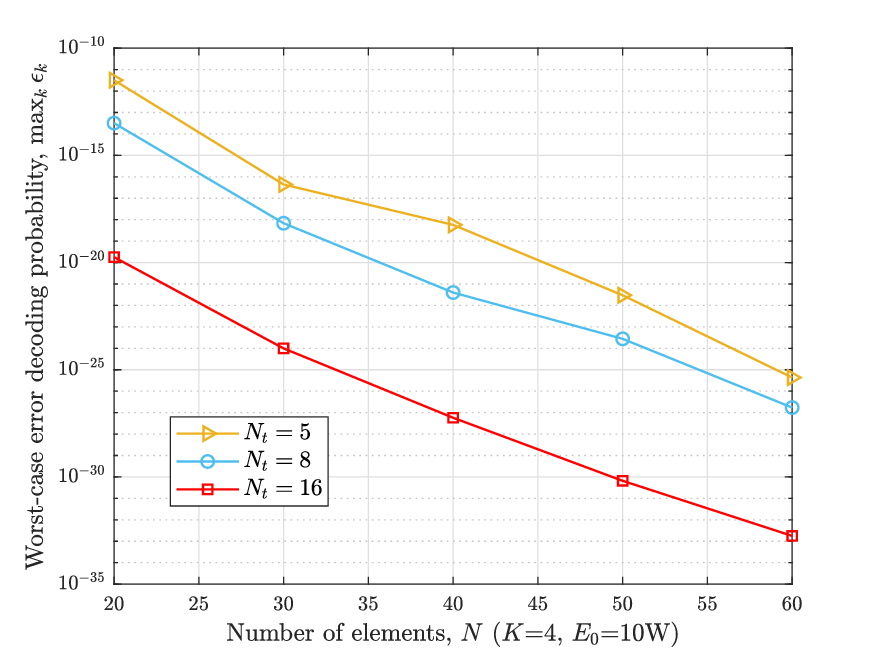}
	\caption{{\color{black}Optimized min-max decoding-error probability versus the number of elements of the RIS when $N_t$ varies.}}
	\label{fig_NTvsN}
\end{figure}

Fig.~\ref{fig_NTvsN} shows the worst decoding error probability when the number of elements in the reconfigurable intelligent surface (RIS), denoted as $N$, is varied across different numbers of base station (BS) antennas, denoted as $N_t$. When $N_t$ increases, there are higher receiving-beamforming gains. That is the reason why the worst-case decoding error goes down when $N_t$ goes up. 
It has also been shown that increasing the number of elements $N$ in the reconfigurable intelligent surface (RIS) has a positive impact on improving the signal-to-interference-plus-noise ratio (SINR) for the worst-case users. Increasing the value of $N$ results in more reflection beamforming gains, hence reducing the min-max decoding error.
 
\begin{figure}[t]
\centering
    \includegraphics[width=0.98\linewidth ]{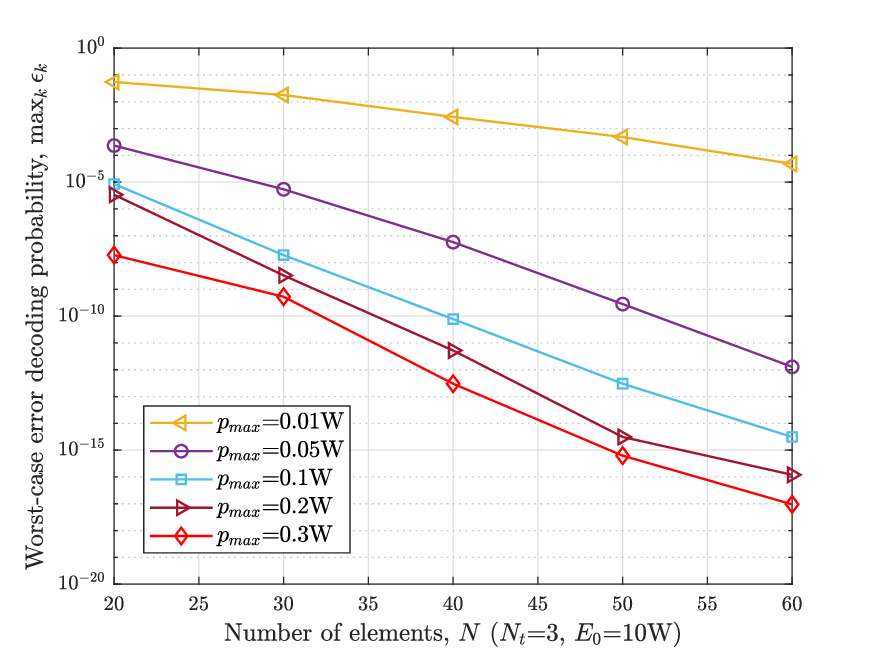}
	\caption{{\color{black}Optimized min-max decoding-error probability versus the number of elements of RIS when $p_{max}$ varies.}}
	\label{fig_E0vsK}
\end{figure}

In Fig. \ref{fig_E0vsK}, we can witness the variation in the worst-case error decoding error as the number of reconfigurable intelligent surface (RIS) elements changes while considering different maximal power levels of the user, denoted as $p_{max}$. It shows that when $p_{max}$ increases, the worst-case decoding error decreases. That is because higher SINR can be obtained by users when $p_{max}$ increases.
\begin{figure}[t]
\centering
    \includegraphics[width=0.98\linewidth ]{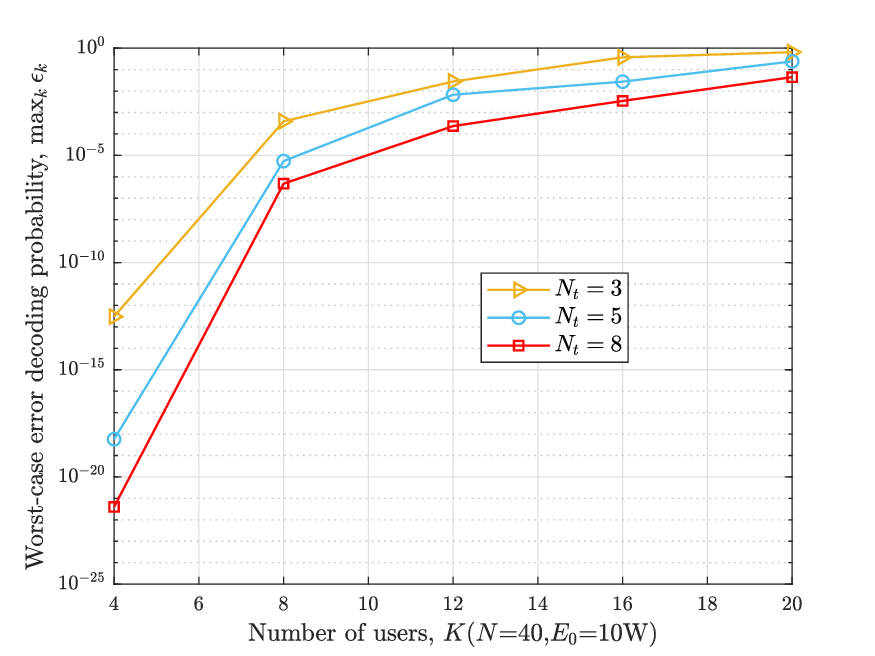}
	\caption{{\color{black}Optimized min-max decoding-error probability versus
a number of users when $N_t$ varies.}}
	\label{fig_EvsK}
\end{figure}
Fig. \ref{fig_EvsK} depicts the variation in worst-case error decoding error as the number of users changes for different numbers of antennas at the base station $N_t$. In addition, Fig. \ref{fig_NTvsK} shows the worst-case error decoding error when the number of users $K$ varies under various total energy limitations $E_0$. Both Fig. \ref{fig_EvsK} and Fig. \ref{fig_NTvsK} show that the worst-case error decoding error increases as the value of $K$ grows. 
\begin{figure}[t]
\centering
    \includegraphics[width=0.98\linewidth ]{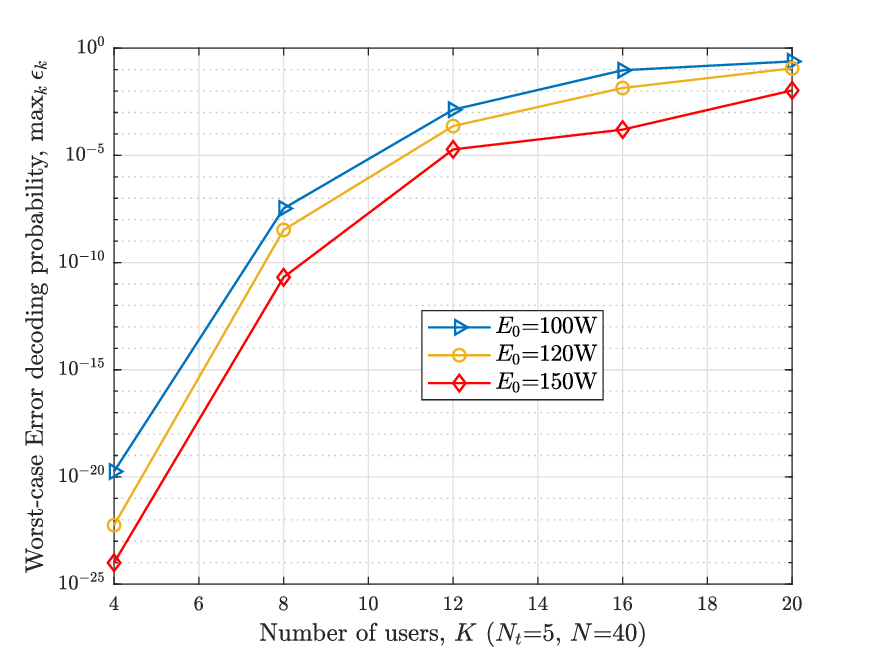}
	\caption{{\color{black}Optimized min-max decoding-error probability versus
a number of users when $E_0$ varies.}}
	\label{fig_NTvsK}
\end{figure}
\section{Conclusion}\label{sec_con}
This paper investigated a RIS-empowered uplink hybrid NOMA-TDMA communication system under short packet transmission. We formulated an optimization problem to minimize the maximum decoding error probability. Then, we proposed an algorithm to jointly optimize the users' transmit power, receiving beamforming vector at the BS, RIS reflection, blocklength, and user pairing. Numerical results have shown that the proposed algorithm can perform better than benchmark schemes, and the proposed system outperforms the pure NOMA system. In future work, we may target other metrics, such as sum rate maximization and average sum of age of information (AoI) \cite{gao2022non}. Since we assume that the channel gain is fixed during the total transmission time in this work, the TDMA time slot duration optimization can be added in future work.


%

\ifCLASSOPTIONcaptionsoff
  \newpage
\fi
\bibliographystyle{IEEEtran}      
\bibliography{AoI}

\end{document}